\documentclass[prl,twocolumn,showpacs,floatfix,superscriptaddress]{revtex4}
\usepackage{graphicx}
\usepackage{amsmath}
\usepackage{latexsym}
\usepackage{dcolumn}

\newcolumntype{f}[1]{D{.}{.}{#1}}

\begin{document}

\title{Low-energy X-ray standards from hydrogenlike pionic atoms}
\author{D.F. Anagnostopoulos}
\affiliation{Department of Materials Science and Engineering, University of Ioannina, 
           GR-45110 Ioannina, Greece}
\author{D. Gotta}
\affiliation{Institut f\"ur Kernphysik, Forschungszentrum J\"ulich, 
           D-52425 J\"ulich, Germany}
\author{P. Indelicato}
\email{paul.indelicato@spectro.jussieu.fr}
\homepage{http://dirac.spectro.jussieu.fr/}
\affiliation{Laboratoire Kastler Brossel,
\'{E}cole Normale Sup\'erieure et Universit\'e Pierre et Marie Curie,
Case 74, 4 place Jussieu, 75005 Paris, France}
\author{L.M. Simons}
\affiliation{Paul-Scherrer-Institut (PSI), CH 5232-Villigen, Switzerland}

\date{September 2, 2003}

\begin{abstract}
We demonstrate the first step of a complete program, which consists in
establishing an X-ray energy standard scale with the use of few-body
atoms, in the few keV range. Light pionic and muonic atoms as well as
one and two-electron ions from Electron-Cyclotron Ion sources are
used.  The transition energies are calculable from quantum
electrodynamics, meaning that only a very limited subset need be
measured and compared with theory, while providing a large number of
standard lines. Here we show that circular transitions in pionic neon
atoms, completely stripped from their electrons, reveal spectral lines
which are narrow, symmetric and well reproducible. We use these lines
for the energy determination of transition energies in complex
electronic systems, like the K$\alpha_{1,2}$ transitions in metallic
Ti, which may serve as secondary standard.
\end{abstract}

\pacs{06.20.Fn, 32.30.Rj, 36.10.-k, 07.85.Nc}
\maketitle


Accurate (below 1~ppm) and \emph{reproducible} X-ray wavelength
standards with reasonably dense set of lines would be very valuable
for the most widespread application of X-rays: the determination of
crystal lattice parameters with diffractometric methods (see e.g.,
Ref.~\cite{hgbw91} and references therein.) Other practical
applications are found, like the energy calibration of synchrotron
radiation beams, monochromators and spectrometers and the
determination of the response function of X-ray spectrometers and
diffractometers. X-ray standards can thus be useful in many areas of modern
science like crystallography, solid state, molecular, atomic and
particle physics, chemistry, and biochemistry.

A recent experiment used the $^{57}$Fe M\"{o}ssbauer radiation,
excited by synchrotron radiation, improved the energy (wavelength)
standard for the energy region of 14~keV by two orders of magnitude in
accuracy from 10~ppm to 0.2~ppm \cite{sljl00}.  This attempt, while
very promising, is very difficult to extend to lower energies, where
electron conversion would dramatically reduce the nuclear
fluorescence. In the absence of an appropriate excitation source, such
as synchroton radiation, this would require unrealistically high
source activities with the additional requirement of a sufficiently
long life time of the parent isotope. In addition self-absorption of
low energy X-rays in the source is very strong. The X-rays can thus
only originate from the surface layer, which leads to an upper bound
to the maximum effective activity that can be reached by increasing
the amount radioactive material. Finally,
all these transitions are orders of magnitude narrower  than crystal spectrometers
resolution.  For some applications the
extreme narrowness of $\gamma$ lines is of no use while limiting
severely their intensity.

The most widely used X-ray energy standards, at present time, are made by exciting inner-shell
transitions in atoms with either electrons or photons. In a number of
cases their energies are given with precision close to 1~ppm
\cite{dkib2003}, which does not necessarily mean that these standards
can be used to such an accuracy. For X-rays originating from
inner-shell transitions in multi-electron systems, the center of
gravity of the line cannot be attributed unambiguously to a physical
transition. Shake-off processes (which create additional vacancies)
and open outer shells lead to numerous satellite transitions very
close in energy to the diagram line, which cannot be resolved and
produce asymmetric line shapes. Moreover, the line shape of
transitions in multielectronic systems depends also on the excitation
mechanism used to create the inner-shell vacancies. For example the
evolution of the K ($1s^{-1}$) Argon spectrum has been studied as
function of excitation energy \cite{dcl82,dlch83}, and dramatic
qualitative changes were observed. The chemical environment of the
atom also plays a strong role as can be seen from the comparison
between solids, metallic vapors and theoretical X-ray absorption edges
energies \cite{dkib2003,ibl98,knnf90}. This
problem also affects transition energies, particularly when they involve M$%
_{2,3}$ and N$_{2,3}$ shells.

An inherent problem with current standard X-ray lines is their natural
width, which is typically more than 10 times larger than the resolution of
the best X-ray spectrometers. Hence, fluorescence radiation is
unsuitable to determine the response function of the apparatus.


As an alternative and more general approach to both $\gamma$-rays or
natural X-rays, we thus propose to profit from recent developments in
exotic-atoms research and in heavy-ions sources, and to use two- and
three-body systems as photon emitters in the few keV range. In
contrast to $\gamma$-rays, electronic, muonic and pionic atoms would
provide a dense set of lines, that can be supplemented by an even
denser set if one can use antiprotonic atoms with beam intensities
comparable to LEAR at CERN \cite{gaab99}.
Our program consists first in doing relative energy
measurements of transitions in one and two-electron ions, emitted by
the plasma of Electron-Cyclotron Ion Sources (ECRIS), of circular
transitions in fully stripped pionic atoms and of X-rays from solid
fluorescence targets. Modern, commercial, permanent-magnet ECRIS are
small and relatively economical to operate, and could be 
available in a large number of places to provide reference lines.

This relative energy scale will then be tied to a few, very bright
lines, the energy of which will be measured absolutely, with either a
double-flat crystal instrument, or a backscattering spectrometer as in
\cite{sljl00} obtained from a Electron-Cyclotron Ion Trap (ECRIT), a
device derived from the ECRIS, and optimized for increasing the
trapping time of the ions \cite{bsh2000}, and thus the production of
X-rays from highly-charged ions. Intense M1 radiation from the
$1s2s\,^1S_0 \to 1s^2$ transition in helium-like Argon has been
observed both in conventional ECRIS \cite{dkgb2000} and in the first
run of the PSI ECRIT in 2002 \cite{abbd2003}.  The ions energy
in such a device ranges from $\approx 0.5$~eV to 6~eV depending on the
injected RF power\cite{ber96}, giving rise to a Doppler broadening in the 5 to 18~ppm,
 i.e., 0.07 to 0.28~eV for Ar.  This allows for
measurements well below 1~ppm, and corresponds also to the expected
accuracy (typically 1~meV) of theory for one electron ions in this range of $Z$.
Different energy ranges and diffraction orders can be connected by
exploiting the Yrast structure of the exotic-atom cascade, which leads
to a strong population of circular states, thus favoring transitions
with a change of the principal quantum number $n$ by 1. Hence,
successive transitions in the same atom connect different energy
scales in ratios $\approx \left( 1/(n+1)^{2}-1/n^{2}\right) /\left(
1/n^{2}-1/(n-1)^{2}\right) $, which are roughly of the order of 0.5
for $n$ around 5. For instance, $\pi$Ne provides the photon energies
of 2.7, 4.5 and 8.3~keV corresponding respectively to the
$7\rightarrow 6$, $6\rightarrow 5$ and $5\rightarrow 4$
transitions. With such a comparison method one can transfer energy
standards between orders without problems due to changes in index of
refraction.

The present program has several advantages. First, once a set of lines
has been measured and compared to a given accuracy to
quantum-electrodynamics (QED) calculations, other lines from the same
source or from neighboring elements can be used as standards without
the need of a direct measurement.  Transition energies of these simple
system can probably be calculated nowadays from first principle to
better than 1~meV, benefiting from high-accuracy tests of QED in
hydrogen \cite{bsaj2000}. This is even true for hadronic atoms, if one
uses circular transitions, which are not affected by strong
interaction.  Only particle masses (very well known except for pion),
the fine structure constant and the Rydberg constant are needed. Over
time the precision of the calculation can be improved by
systematically including contributions from higher-order Feynman
diagrams.  The quality of the calculation can be checked by comparing
to the directly measured lines energies as well as to all the
relatively measured ones which are directly connected to the direct
measurements.

Second, the natural line widths of these transitions, while not as
small as $\gamma$-ray line, are three orders of magnitude narrower
than the ones of multielectronic systems.

Finally, the combination of the exotic-atom and electronic
X-ray sources can provide a powerful tool for establishing
high-quality X-ray energy standards. As an example, the $5g\rightarrow 4f$
transition in $\pi $C, the Lyman-$\alpha$ in hydrogenlike Cl and the K$\alpha$
fluorescence line in singly ionized Ar, degenerate in energy by few
eV, can be easily related by the method presented here.

In the present letter, we present the measurement, with a crystal
spectrometer of the characteristic X-radiation from hydrogenlike
pionic atoms. We use these transitions as energy standards for the
energy determination of the transition energies in a complex
electronic systems, the K$\alpha _{1,2}$ transitions in Ti.  In this
way we intend to show that the pionic transition is in good agreement
with a reasonably well measured K$\alpha _{1,2}$ transition. Copper
would have made a better case, but the present world average for the
pion mass involve the Cu K$\alpha$ doublet. This issue will however be
solved when the final value for the pion to muon mass ratio is
released \cite{naab2002}.

For this experiment we used the cyclotron trap II \cite{sim93} attached to
the $\pi $E5 pion line at the Paul Scherrer Institute (PSI, Switzerland). In
this device a 112~MeV/c pion beam is decelerated in a magnetic field using a
suitable set of degrader foils. Such a set-up allows the use of dilute
targets like gases. Typically $4\times 10^{8}$~$\pi ^{-}$/s are injected in
the trap for 1~mA proton current. The target consists of a cylinder of 60~mm
diameter and 26~cm length, with 50~$\mu $m-thick kapton walls. The pressure
in the target was around 1~bar, leading to typically $1.7\times 10^{6}$~$\pi
^{-}$/s stops in the gas.

For light elements ($Z\leq 10$) the cascade that follows leads quickly to
the formation of an hydrogenlike exotic atom in a circular state. All the
electrons are ejected by Auger effect in the early stage of the cascade, in
a process similar to internal conversion in nuclei, because of the large
mass of the pion ($\approx 273\times m_{e^{-}}$). Accordingly transition
energies are 273 times larger than electronic ones between states of
identical quantum numbers. Because the atoms are formed in a low-pressure
gas the time it takes to recapture electrons from molecules in the gas is
much longer than the pionic atom lifetime, hence, the exotic
atoms stay in an hydrogenlike state for the rest of the atomic cascade. It
has been shown that less than 2~\% of the X-ray observed are affected by the
presence of an extra electron \cite{lbgg98}. When using solid targets the
undefined status of the electron shell is the principal limitation in
high-precision experiments using exotic-atom X-rays \cite{jgl94}.

X-rays emitted by the exotic atoms at the center of the trap are analyzed
using a Johann-type Bragg spectrometer that was developed for applications requiring
very high luminosity and excellent resolution in noisy environments, in the
1.7 to 10~keV range. It was equipped with a spherically bent \ Si(220)
crystal having a radius of curvature of 2.9854~m. A detailed description of the
apparatus is given in ~\cite{lbgg98, gaab99, asgd99} .

\begin{figure}[tbp]
\centering
\includegraphics[height=8cm]{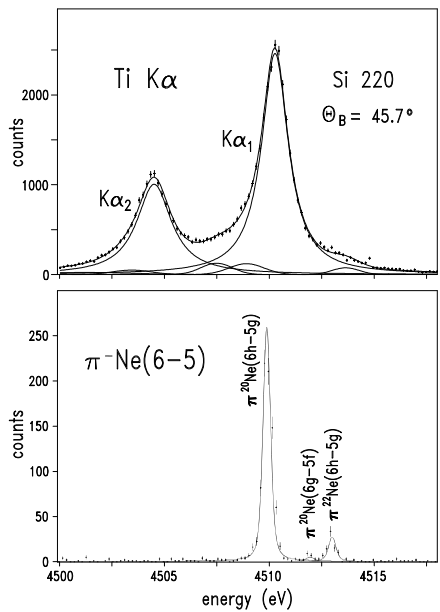}
\caption{Upper : Ti K$a$ doublet fitted by a sum of six Voigt functions.
Lower : $\protect\pi $Ne $6\rightarrow 5$ transitions showing fine-structure
and isotopic effects. Comparison of the two spectra demonstrates the energy
coincidence of the Ti K$\protect\alpha _{1}$ and of the $\protect\pi ^{20}$%
Ne $6h\rightarrow 5g$ lines.}
\label{fig:pine-ti}
\end{figure}

For this proof-of-principle measurement \ we relate the energy of the $%
6h\rightarrow 5g$ transition in pionic neon to energy of the K$\alpha _{1,2}$ transitions
of Ti. The Ti K$\alpha _{1}$ differs only by a fraction
of an eV from the strong $6h\rightarrow 5g$ transition in $\pi ^{20}$Ne.
Hence, the measurement could be performed by exchanging only the neon-filled
target cell with a $30\times 20$~mm$^{2}$ plate of metallic Ti, without any
other change of the experimental set-up. The fluorescence X-rays were
excited by means of an X-ray tube with a Cr-anode. The consecutively
recorded $\pi $Ne and Ti spectra are shown in Fig.~\ref{fig:pine-ti}.

The pionic transition energies are calculated from the world average
pion mass $m_{\pi }=139.57018\pm 0.00035$~MeV as given by the particle
data group \cite{PDBook} and fundamental constants \cite{mat2000} with
the Klein-Gordon equation for a spherical nuclear charge distribution
(to improve numerical stability although direct effect on energy is
small). They include the Uehling potential for vacuum polarization to
all order, the K\"{a}ll\`{e}n and Sabry as well as the Wichman and
Kroll correction, and include nuclear recoil and relativistic
recoil. The nuclear masses for $^{20} $Ne and $^{22}$ Ne are
deduced from atomic masses in \cite{dnbp94} and \cite{aaw95}
respectively, by subtracting the mass of the 10 missing
electrons. These energies can be calculated with high precision since
the strong interaction plays no role for such high-lying circular
states. We obtain 4509.894~eV and 4512.948~eV for the $6h\rightarrow
5g$ transition in $\pi ^{22}$Ne and $\pi ^{20}$Ne, respectively. There
is an uncertainty of 11 meV on these energy values which originates
exclusively from the pion mass.

The natural width of the $\pi$Ne $6h\rightarrow 5g$ transition is 12~meV.
The fact that the calibration line is almost a $\delta $ function allows for
precise determination of the spectrometer response function. The measured
instrumental response function width of 440$\pm$20~meV is close to the theoretical limit of 330~meV as
predicted from the Monte-Carlo simulations for the chosen geometry. The $\pi 
$Ne line shape is fitted sufficiently well by using Gaussian line profiles.

To deduce the Ti K$\alpha $ line shape we fitted a sum of six Voigt
profiles, following \cite{asgd99} using the Gaussian response function
obtained from the $\pi $Ne spectrum. The peak positions, K$\alpha
_{1,\,2}^{0}$, are obtained from the zeros of the derivative of the
fitted function. The spectrometer dispersion, necessary to transform
the position information of the detector to energy, is obtained in a
self-consistent way from the $\pi $Ne spectrum itself (Fig.~ \ref{fig:pine-ti}).
For the very small energy difference, for the case of $\pi $Ne $%
6\rightarrow 5$ and Ti K$\alpha $, the energy-dependent corrections
originating from the imaging properties of crystal spectrometer
(refraction index, crystal bending and size, rocking curve ) are
almost invariable and cancel out.

The results are displayed in Table \ref{tbl:results}, together with
presently known values. The experimental error in the energy of the K$\alpha
_{1}$ line is practically given by the statistical uncertainty on the K$%
\alpha _{1}$ and $\pi ^{20}$Ne $6h\rightarrow 5g$ lines. In the case of the K%
$\alpha _{2}$ line the error is dominated by the uncertainty on the
dispersion value, which is due to the limited statistics of the $\pi ^{22}$%
Ne $6h\rightarrow 5g$ transition. Both contributions could be
drastically reduced by increasing the statistics of the measurement.

\begin{table*}[tbp]
\caption{Peak position energies and natural widths of the K$\protect\alpha %
_{1,2}$ transitions in metallic Ti in comparison with previous
measurements (in eV). For this work the errors on the energies from the
experiment (first parenthesis) and from the calibration standard, i. e., the
uncertainty of the pion mass (second parenthesis), are given separately.}
\label{tbl:results}\centering
\begin{ruledtabular}
    \begin{tabular}{ccdddd}
Element &  Line    &\multicolumn{1}{c}{Energy (this work)}     &\multicolumn{1}{c}{Energy (Refs.~\cite{dkib2003})}               &
\multicolumn{1}{c}{Width (this work)}       & \multicolumn{1}{c}{Width}     \\
\colrule                                        
\colrule                                                      
Ti  &   K$\alpha_1^0$ &   4510.903    (19)  (11)  &   4510.869    (49) &   1.6   (1)   &   1.5 (3)  \footnote{Ref.~\cite{sal76}}\\
    &   K$\alpha_2^0$ &   4504.942    (40)  (11)  &   4504.887    (49) &   2.1   (1)   &   2.1 (4) \footnotemark[1] \\
  \end{tabular}
    \end{ruledtabular}
\end{table*}

 Narrow transitions from exotic atoms allow to characterize very precisely
the response function of a curved crystal set-up.  For the described
experiment the accuracy of the extracted natural widths of the electronic
systems (Table \ref{tbl:results}) reaches the one obtained with ultimate
resolution devices like double-flat crystal spectrometers  \cite{knnf90, sal76}.

In the present letter we have demonstrated that narrow lines from
hydrogenlike pionic atoms are able to serve as energy standards in the
few keV range. By using this method the energy uncertainty is limited
primarily by the knowledge of the charged pion mass, whenever
transitions close in energy can be found. We have proven that an
hydrogenic pionic line, the energy of which has been calculated from
QED can be used to establish the energy of a previously well measured line with
comparable accuracy. A research program is underway to improve on the
precision of the pion mass to the order of about 1~ppm
\cite{naab2002}. \ An alternative is to use muonic transitions as the
muon mass is known to 0.05~ppm \cite{PDBook}. However, intensities
achievable are about two orders of magnitude smaller than in the case
of pionic atoms. Much higher intensity will be available from X-ray
sources like the super-conducting Electron-Cyclotron Ion Trap (ECRIT)
developed at the Paul Scherrer Institute, in which few-electron atoms
up to hydrogenlike systems are produced \cite{bsh2000}. Such
electronic two-body systems will be used as calculable energy
standards in the same way as exotic atoms.

In a separate experiment we have recently used the technique presented
here to measure the energy of the Sc K$\alpha$  lines, which are known only
from interpolation \cite{bea67}, with an accuracy improved by a factor of 12
 \cite{agis2003}.

By using X-ray lines, both from hydrogenlike exotic and electronic
atoms a relative energy scale is established that can easily range up
to 30 keV. Around 60 lines will be available if electronic, pionic and
muonic atoms are used, hundred more if antiprotonic atoms are
available. The energy of these lines will depend only on the fine
structure constant $\alpha $, and on the pion, muon and electron mass,
and they can also be connected to low-energy $\gamma $-ray standards.

Laboratoire Kastler Brossel is Unit\'e Mixte de Recherche du CNRS n$^{\circ}$
8552, of the Physics Department of \'Ecole Normale Sup\'erieure and
Universit\'e Pierre et Marie Curie.


\end{document}